\documentstyle[aps,prl,epsfig]{revtex}
\draft
\newcommand{\size}{10}
\newcommand{\sizebig}{20}
\begin{document}
\title{Quasibound states at thresholds in multichannel impurity scattering}
\author{Sang Wook Kim, Hwa-Kyun Park, H.-S. Sim, and Henning Schomerus}
\address{Max-Planck-Institut f{\"u}r Physik komplexer Systeme,
N{\"o}thnitzer Str. 38, D-01187 Dresden, Germany}
\date{\today}

\maketitle

\begin{abstract}

We investigate the threshold behavior of transmission resonances and quasibound states 
in the multichannel scattering problems of a one dimensional (1D) time-dependent impurity 
potential, and the related problem of a single impurity in a quasi 1D wire. It was 
claimed before in the literature that a quasibound state disappears when a transmission 
zero collides with the subband boundary. However, the transmission line shape, the Friedel 
sum rule, and the delay time show that the quasibound states still survive and affect the 
physical quantities. We discuss the relation between threshold behavior of transmission 
resonances, and quasibound states and their boundary conditions in the general context of 
multichannel scatterings. 

\end{abstract}



\section{introduction}

Electronic transport properties in quasi one-dimensional (1D) wires have gained much 
attention, not only due to scientific interest but also due to practical applications
\cite{Datta95}. When an electron is restricted to a wire, confinement subbands are 
formed due to quantization of the transverse momentum. In clean straight wires without 
impurities the modes of the subbands decouple, which results in a perfectly quantized 
conductance. However, the modes (including evanescent ones) mix when impurities are 
introduced into the wire, and the transport properties show a rich and non-trivial 
behavior. Detailed aspects of the mixing have been extensively investigated 
\cite{Datta,Bagwell90,Bagwell92,Levinson92,Kunze92,Wang97}.
The impurities put into the wire usually give rise to bound states, which can be 
separated into two categories: true bound states with real energy and {\em quasibound} 
states with complex energy. The imaginary part determines the decay rate of the 
quasibound state.

Even in simple 1D scattering problems quasibound states are quite common, for example 
resonant transmission states through a double barrier \cite{Price88,Buettiker88} and 
Ramsauer-like resonances \cite{Ramsauer21} in potential wells. According to the 
Breit-Wigner theory \cite{Breit36}, for each quasibound state the transmission amplitude 
possesses a pole in the complex energy plane. In multichannel scattering problems another 
kind of quasibound states can appear; 
the Fano type resonance originally proposed to explain autoionization in 
atomic physics \cite{Fano61}. When a discrete energy level interferes with a continuum of 
states, the excitation spectra become asymmetric. Each subband mode in a quasi-1D wire
gives rise to a zero-pole pair of transmission in the complex energy plane, where the 
zero occurs on the real energy axis. As a consequence, there exist transmission 
zeros, and the zero-pole pairs lead to asymmetric resonance features \cite{Porod93}.

Transport through a time-dependent (mostly periodically oscillating) potential is also 
a subject of increasing importance, with a growing number of applications
\cite{Buettiker82,Wagner,Sun98,Burmeister,Switkes99,Park00,Martinez01,Henseler01}.  
One of the important features here is that an oscillating potential can transfer an incoming 
electron of energy $E$ with finite probability to `Floquet' sidebands at $E\pm n\hbar 
\omega$, where $n$ is an integer and $\omega$ is the angular frequency of the oscillation. 
Even though this corresponds to an inelastic scattering, the process is coherent and 
non-dissipative. Hence, it can still be regarded as a multichannel scattering problem 
like in the quasi-1D case, with the sidebands being analogous 
to the subbands. However, from the physical point of view, the Fano type quasibound states of 
the periodically oscillating impurity are somewhat different from those of quasi-1D system 
\cite{Li99}. If a bound state exists due to an attractive static potential added to the 
oscillating potential, electrons in 
the incident channel can emit photons and drop to the bound state, and similarly electrons 
in the bound state can absorb photons and jump to the incident channel. A transmission 
resonance takes place when the energy difference between the incident channel and the bound 
state is equal to the energy of one or more photons of the oscillating potential. Even 
without bound states transmission resonances like these can occur.

The multichannel subbands in the quasi-1D case and the Floquet sidebands in the oscillating 
potential case divide the complex energy plane into sub-domains with appropriate boundary 
conditions for the quasibound states. Upon varying an external parameter such as the strength 
of the impurity potential, the quasibound states change their location in complex energy plane. 
In the literature \cite{Martinez01} it was claimed for the oscillating impurity that the 
quasibound state vanishes when the transmission zero collides with the sub-domain boundary
and disappears.  However, as we will demonstrate, the transmission line shape, the Friedel 
sum rule, and the delay time show that 
{\em the quasibound states do not disappear abruptly}. This clarifies the relation between 
threshold behavior of transmission resonances, and quasibound states and their boundary 
conditions in the general context of multichannel scatterings. 

In Sec.~II, we briefly summarize our findings about the transmission resonances and quasibound 
states. In Sec.~III, we investigate the characteristics of quasibound states in the scattering 
problem with a 1D oscillating delta-function impurity, and discuss Friedel sum rule and Wigner
delay time. In Sec.~IV, we study our second example, the scattering from a static delta-function 
impurity in a quasi-1D wire. Finally, we conclude our paper in Sec.~V.


\section{boundary conditions of quasibound states in multichannel scattering}

In this section we summarize the main finding of this paper, which will be further illustrated
in Sec.~III and IV. The transmission coefficient $t_{mn}(E)$ relates the incoming wave 
$\psi^{in}_n$ to the scattered 
wave 
$\psi^{sc}_m$ by
\begin{equation}
\psi^{sc}_m(E) = t_{mn}(E) \psi^{in}_n.
\end{equation}
The resonance energies are found by locating the poles of $t_{mn}(E)$, namely
\begin{equation}
\frac{1}{t_{mn}(E_R-iE_I)} = 0.
\end{equation}
The poles of $t_{mn}(E)$ occur at complex energies $E=E_R-iE_I$. At these complex energies, 
a scattered wave can be produced even if no incident wave is present. When we choose the sign 
convention of time as $\exp(-i E t/\hbar)$, $E_I$ must be positive 
for the system to be stable. In the Breit-Wigner case \cite{Breit36} the transmission near 
the quasibound state is given by $t(E) \sim 1/[E-(E_R-iE_I)]$, where the pole in complex
energy plane completely determines the transmission line shape.

The poles can also be found by solving the wave equation with the appropriate boundary 
conditions of only outgoing waves. This can be described by the following 
linear equation
\begin{equation}
M \left( 
\begin{array}{c}
\psi^l_{out} \\
\psi^r_{out}
\end{array}
\right)=0.
\label{null_eq}
\end{equation} 
Here $M$ is the matrix of wave-matching conditions, and $\psi^l_{out}$ and $\psi^r_{out}$ are 
the left and the right outgoing wave vectors, respectively. The matrix $M$ contains the wave 
number $k$ as a function of energy $E$. For real $E$ the wave number $k$ is given by 
$k=\sqrt{2\mu E}/\hbar$, where $\mu$ is the mass. For complex energies we have to choose the 
sign of the square root carefully. Considering the general relation \cite{Ahlfors79}
\begin{equation}
\sqrt{\alpha+i\beta} = \pm \left(
\sqrt{\frac{\alpha+\sqrt{\alpha^2+\beta^2}}{2}}
+i\frac{\beta}{|\beta|}\sqrt{\frac{-\alpha+\sqrt{\alpha^2+\beta^2}}{2}}
\right)
\end{equation}
provided that $\beta \neq 0$, we obtain
\begin{equation}
\sqrt{E_R-iE_I} = \pm (k_R - ik_I),
\label{sign}
\end{equation}
where by definition $E_I$, $k_R$ and $k_I$ are positive. The real energy $E_R > 0$ for
propagating states, and $E_R < 0$ for evanescent states. The correct choice of the sign in 
Eq.~(\ref{sign}) depends on the analytical continuation based on the physical constraints on 
the real energy axis. For $E_I=0$ the following relation should be satisfied:
\begin{eqnarray}
\sqrt{E_R} = \left\{
\begin{array}{c}
k_R, ~~~ E_R>0 \\
ik_I, ~~~ E_R<0,
\end{array} \right.
\label{real_condition}
\end{eqnarray}
where both $k_R$ and $k_I$ are still positive. The choice of sign for propagating states
corresponds to the choice of the {\em negative} sign for $k_I$ (i.e., the positive sign for 
$k_R$), and follows from the propagating direction of the plane wave far from the scatterer. 
For evanescent states the {\em positive} sign for $k_I$ must be chosen in order to avoid 
divergence of the solution in space. 

It is worth noting that the `$+$' sign corresponds to the decaying wave 
$\exp(ik_Rx-k_Ix)\exp(-iE_Rt/\hbar-E_It/\hbar$) while the `$-$' sign corresponds to the 
diverging wave $\exp(ik_Rx+k_Ix)\exp(-iE_Rt/\hbar-E_It/\hbar$), with $x>0$. The physical 
reason of the exponential divergence of the channels with positive real energy is a retardation 
effect; the wave at $x\gg1$ has propagated away from the impurity where it originated at a time 
$\Delta t \approx x/v$ in the past, where $v$ is the velocity of the wave. However, at that 
earlier time the amplitude of the wave at the impurity was larger by a factor 
$\exp(E_I\Delta t/\hbar)$. This corresponds to $\exp(k_Ix)$ 
since $(E_R-iE_I)/\hbar \simeq v(k_R-ik_I)/2$ \cite{Noeckel97}.

For multichannel scattering the wave numbers are given by
\begin{equation}
k_n = \pm\sqrt{E_R - E_n -iE_I},
\label{sqrt}
\end{equation}
where $E_n$ represents the quantized energy of a transverse mode in the quasi-1D case 
($n = 1, 2, 3, \cdots$) and is equal to $n\hbar\omega$ in the 1D oscillating potential 
case ($n = \cdots, -1, 0, 1, \cdots$). We order the $E_n$'s by magnitude and denote by 
$n^*$ the special value of $n$ which satisfies $E_{n^*} < E_R < E_{n^*+1}$.

In order to satisfy the physical conditions for the signs of $k_n$'s mentioned above, for 
$n \leq n^*$ the `$-$' sign should be chosen for Im$k_n$, and for $n > n^*$ the `$+$' sign, 
which corresponds to the selection of a certain Riemann sheet for evaluating the complex 
square root function. This choice of the signs will be called {\em proper Riemann sheet} 
since this set of signs corresponds to the correct analytical continuation from the real 
energy axis. In order to represent the Riemann sheet through the signs of Im$k$ like in our 
convention, the square root functions should have their branch cuts on the real axis starting
from the branch points and extending to the positive infinity. In general the notion Riemann 
sheet depends on the choice of a branch cut. However, this choice of the branch cut cannot 
affect physical results, because it is only a way to label the projection of the Riemann
sheets onto the complex plane.

\begin{figure}
\center
\includegraphics[height=\sizebig cm,angle=0]{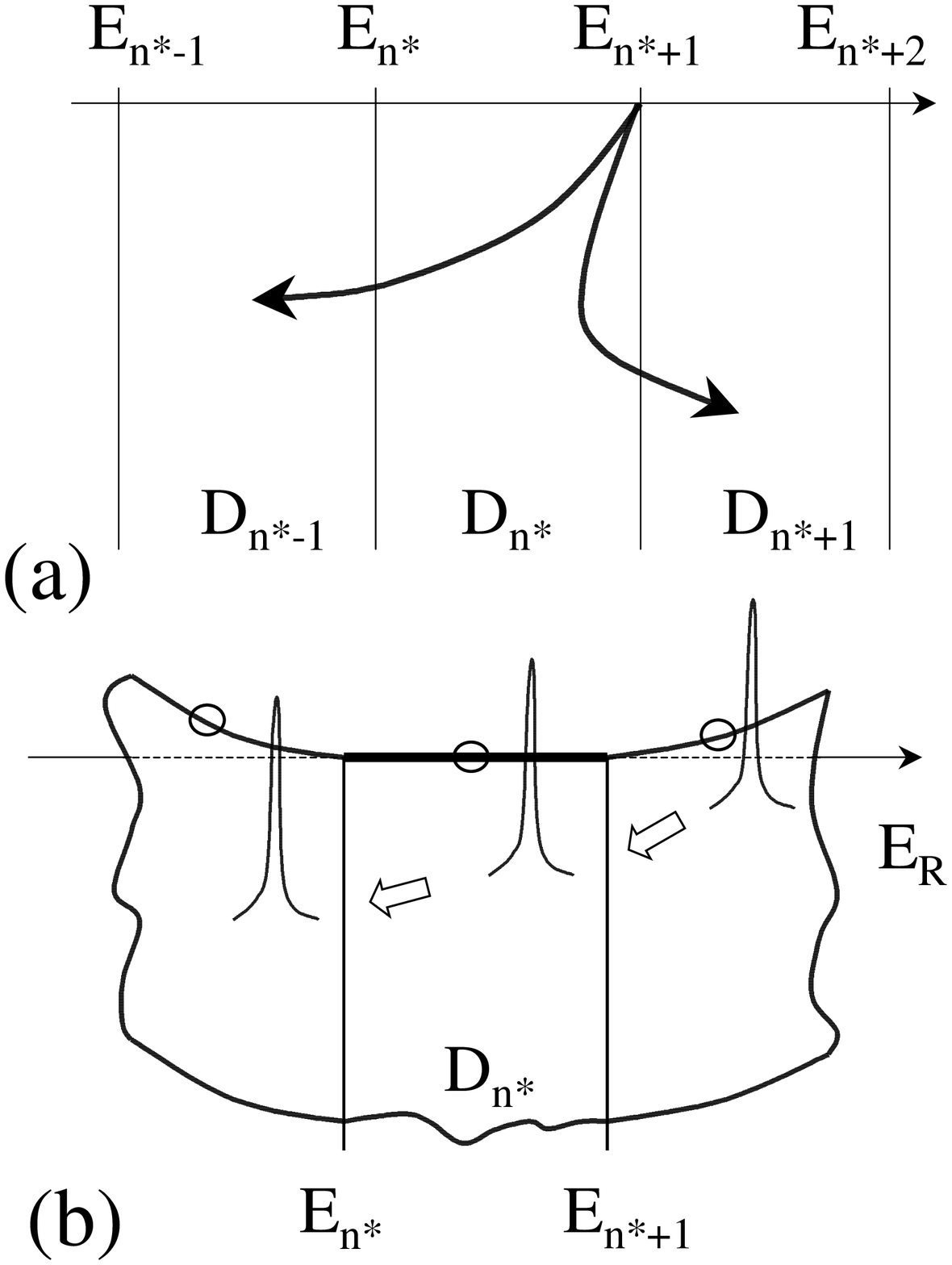}
\caption{(a) Schematic diagram for possible motion of quasibound states in the complex 
energy plane. $E_n$ and $D_n$ denote subband threshold energies and domains, respectively.
(b) Schematic of the proper Riemann surface $R$ of the domain $D_{n^*}$ in complex energy 
plane. The circles and the peaks represent the transmission zeros and the quasibound 
states (or poles), respectively. }
\label{fig1}
\end{figure}

We introduce a labeling for the Riemann sheets by the vector of signs of the square 
root function. In this notation the proper Riemann sheet is represented by 
($\cdots, -, -, \ominus, +, +, \cdots$), which implies `$-$' for Im$k_{n^*-2}$, `$-$' for 
Im$k_{n^*-1}$, `$-$' for Im$k_{n^*}$, `$+$' for Im$k_{n^*+1}$, and `$+$' for Im$k_{n^*+2}$, 
and $\ominus$ has been used to indicate the sign of $n^*$. We also call the regions of 
complex energies separated by the quantized energies $E_n$'s ``domains'', e.g. $D_n$ 
represents the region with $E_n < E_R < E_{n+1}$ as shown in Fig.~\ref{fig1}(a).

As a parameter of the scattering potential such as its strength is varied, the locations 
of the poles change continuously and sometimes can collide with the boundary of the initial 
domain. We can trace the pole into the neighboring domain through the boundary, for example 
from $D_{n^*}$ to $D_{n^*-1}$ in Fig.~\ref{fig1}(a). In fact the trajectory of the pole 
across the boundary is continuous and smooth once we use the same Riemann sheet as the 
used one in $D_{n^*}$, i.e. ($\cdots, -, -, \ominus, +, +, \cdots$). However, this does not 
correspond to the proper Riemann sheet ($\cdots, -, -, \oplus, +, +, \cdots$) in this new 
domain $D_{n^*-1}$ since the inequality $E_{n^*} < E_R$ is violated. Thus, the solution 
obtained in $D_{n^*-1}$ using the proper Riemann sheet of $D_{n^*}$ does not have any 
physical meaning in the domain $D_{n^*-1}$ (more precisely {\em on the real axis of 
$D_{n^*-1}$}), but still can influence physics on the real axis of $D_{n^*}$, especially 
when the resonance is broad. 
Let us emphasize that only the branch points, $E_n$'s, play an important role in the 
threshold behavior of transmission resonances, but not the placement of the cuts. Physics
is in the choice of signs on the real energy axis, Eq.~(\ref{sign}). This situation is 
summarized schematically in Fig.~\ref{fig1}(b), which shows that only small part of the 
proper Riemann sheet of $D_{n^*}$, i.e. $E_{n^*} < E_R < E_{n^*+1}$, 
is attached to the real energy axis. Since in either $D_{n^*-1}$ or $D_{n^*+1}$ 
the quasibound state can survive on the proper Riemann sheet of $D_{n^*}$, it can still 
affect the physics in the region $E_{n^*} < E_R < E_{n^*+1}$, but has no effect on other 
parts of real axis. In the following sections we discuss two examples.


\section{1D delta-function impurity with time dependence}

As the first example we investigate a 1D scattering problem with a delta-function impurity 
oscillating at frequency $\omega$. This problem was studied before by Bagwell and Lake 
\cite{Bagwell92}, who calculated electron transmission and discussed transmission resonances. 
Recently Martinez and Reichl \cite{Martinez01} have investigated this problem and found the 
existence of nonresonant bands in the transmission amplitude as a function of the strength of 
the potential and driving frequency. They observed a periodic behavior of the conductance 
as a function of the scattering strength and of the oscillation frequency of the scatterer.


\subsection{Scattering matrix formulation}

The system is described by the Hamiltonian
\begin{equation}
H(x,t) = -\frac{\hbar^2}{2\mu}\frac{d^2}{dx^2} + [V_s + V_d \cos(\omega t)]\delta(x),
\end{equation}
where the $\mu$ is the mass of an incident particle, and $V_s$ and $V_d$ represent the 
strength of the static and the oscillating potential, respectively. Using the Floquet 
formalism \cite{Reichl92} the solution of this Hamiltonian can be expressed as
\begin{equation}
\Psi_{E_{Fl}}(x,t) = e^{-i E_{Fl} t/\hbar}\sum_{n=-\infty}^{\infty} \psi_n (x) 
e^{-in\omega t},
\label{floquet}
\end{equation}
where $E_{Fl}$ is the Floquet energy which take continuous values in the interval 
$0 < E_{Fl} \leq \hbar \omega$. 

Since the potential is zero everywhere except at $x=0$, $\psi_n(x)$ is given by the 
following form
\begin{equation}
\psi_n (x) = \left\{
\begin{array}{c}
A_n e^{ik_n x} + B_n e^{-ik_n x}, ~~~ x<0 \\
C_n e^{ik_n x} + D_n e^{-ik_n x}, ~~~ x>0,
\end{array} \right.
\label{plane_wave}
\end{equation}
where $k_n=\sqrt{2\mu(E_{Fl} + n\hbar\omega)}/\hbar$.
The wave function 
$\Psi_{E_{Fl}}(x,t)$ is continuous at $x=0$, 
\begin{equation}
A_n + B_n = C_n + D_n,
\label{eq1.1}
\label{osc_bc1}
\end{equation}
and the derivative jumps by
\begin{equation}
\left.\frac{d\Psi_{E_{Fl}}}{dx}\right|_{x=0^+} - \left.\frac{d\Psi_{E_{Fl}}}{dx}\right|_{x=0^-} 
=\frac{2m}{\hbar^2}[V_s + V_d \cos(\omega t)]\Psi_{E_{Fl}}(0,t).
\label{osc_bc2}
\end{equation}
Using Eq.~(\ref{floquet}) this leads to the condition
\begin{eqnarray}
ik_n (C_n - D_n - A_n + B_n) 
& = & \gamma_s(A_n+B_n) +\gamma_d(A_{n+1}+A_{n-1}+B_{n+1}+B_{n-1}) 
\label{eq2.1}\\
& = & \gamma_s(C_n+D_n) +\gamma_d(C_{n+1}+C_{n-1}+D_{n+1}+D_{n-1})
\label{eq2.2}
,\end{eqnarray}
where $\gamma_s=2\mu V_s/\hbar^2$ and $\gamma_d=\mu V_d/\hbar^2$. After some algebra
we have the following equation from Eqs.~(\ref{eq1.1}), (\ref{eq2.1}), and (\ref{eq2.2})
\begin{equation}
\left(
	\begin{array}{c}
	\vec{B} \\ \vec{C}
	\end{array}
\right)
=
\left(
	\begin{array}{cc}
	-(I+\Gamma)^{-1}\Gamma & (I+\Gamma)^{-1} \\
	(I+\Gamma)^{-1} & -(I+\Gamma)^{-1}\Gamma
	\end{array}
\right)
\left(
	\begin{array}{c}
	\vec{A} \\ \vec{D}
	\end{array}
\right)
\label{mat_eq}
,\end{equation}
where 
\begin{equation}
\Gamma = 
\left(
	\begin{array}{ccccc}
	\ddots & \ddots & 0 & 0 & 0 \\
	\gamma_d/ik_{-1} & \gamma_s/ik_{-1} & \gamma_d/ik_{-1} & 0 & 0 \\
	0 & \gamma_d/ik_0 & \gamma_s/ik_0 & \gamma_d/ik_0 & 0 \\
	0 & 0 & \gamma_d/ik_1 & \gamma_s/ik_1 & \gamma_d/ik_1\\
	0 & 0 & 0 & \ddots & \ddots
	\end{array}
\right),
\end{equation}
and $I$ is an infinite-dimensional square identity matrix. Eq.~(\ref{mat_eq}) can also be 
expressed in the 
form $\left.|{\rm out}\right> = M \left.|{\rm in}\right>$, where $M$ connects the input 
coefficients to the output coefficients including the associated evanescent Floquet 
sidebands. In order to construct the scattering matrix we multiply an identity to both sides, 
$K^{-1}K\left.|{\rm out}\right> = M K^{-1}K\left.|{\rm in}\right>$, where $K_{nm}=\sqrt{k_n}
\delta_{nm}$. Then we have $\vec{J}_{out}=\bar{M}\vec{J}_{in}$, where $\vec{J}$ represents
the amplitude of probability flux and $\bar{M} \equiv KMK^{-1}$. It should be mentioned 
that $\bar{M}$ is not unitary due to the evanescent modes included.

If we keep only the propagating modes, we obtain the unitary scattering matrix $S$
\cite{Henseler01,Li99}, which can be expressed in the following form
\begin{equation}
S = \left(
	\begin{array}{cccccc}
	r_{00} & r_{01} & \cdots & t'_{00} & t'_{01} & \cdots \\
	r_{10} & r_{11} & \cdots & t'_{10} & t'_{11} & \cdots \\
	\vdots & \vdots & \ddots & \vdots  & \vdots  & \ddots \\
	t_{00} & t_{01} & \cdots & r'_{00} & r'_{01} & \cdots \\
	t_{10} & t_{11} & \cdots & r'_{10} & r'_{11} & \cdots \\
	\vdots & \vdots & \ddots & \vdots  & \vdots  & \ddots \\
	\end{array}
\right),
\label{osc_smatrix}
\end{equation}
where $r_{nm}$ and $t_{nm}$ are the reflection and the transmission amplitudes, respectively,
for modes incident from the left; $r'_{nm}$ and $t'_{nm}$ are similar quantities for modes 
incident from the right. From $S$ we can obtain the total transmission coefficient for the
propagating mode entering in the $m$th channel
\begin{equation}
T_m(E_{Fl})=\sum_{n=0}^{\infty}\left|t_{nm}\right|^2.
\end{equation}

Figure \ref{fig2} shows the transmission coefficient $T$ as a function of $\varepsilon$
for various values of a dimensionless potential strength $a_d$, where 
$\varepsilon=E/\hbar\omega$ ($E$ is the kinetic energy of the incident particle) 
and $a_d=\mu V_d^2/8\hbar^3\omega$. We exploit the relation 
$T(E)=T_m(E_{Fl})$ where $E=E_{Fl}+m\hbar\omega$. For small $a_d$'s the transmission coefficient 
shows some signatures of Fano type transmission resonance structures [Fig.~\ref{fig2}(a) 
and (b)] whereas the resonance structure and transmission zero disappear as we increase 
$a_d$ [Fig.~\ref{fig2}(c)]. Although there exists no transmission zero in Fig.~\ref{fig2}(c),
the resonance-like structure (the peak around $\varepsilon=0.8$) still remains. For much 
larger $a_d$ the transmission zero reappears from $\varepsilon=1$. Let us note that
the transmission coefficient has no symmetry of energy translation by $m\hbar\omega$
since in general $T_m(E_{Fl}) \neq T_{m'}(E_{Fl})$ when $m\neq m'$.


\subsection{Quasibound states}

In order to obtain quasibound states we apply the same procedure used in the previous
subsection, but with the different boundary conditions, $A=0$ and $D=0$. Since $B_n=C_n$
because of the reflection symmetry at $x=0$, after little algebra we obtain the linear 
equation
\begin{equation}
\left(
	\begin{array}{ccccc}
	\ddots & \ddots & 0 & 0 & 0 \\
	i\sqrt{a_d} & i\sqrt{a_s}+\sqrt{\varepsilon-1} & i\sqrt{a_d} & 0 & 0 \\
	0 & i\sqrt{a_d} & i\sqrt{a_s}+\sqrt{\varepsilon} & i\sqrt{a_d} & 0 \\
	0 & 0 & i\sqrt{a_d} & i\sqrt{a_s}+\sqrt{\varepsilon+1}  & i\sqrt{a_d} \\
	0 & 0 & 0 & \ddots & \ddots
	\end{array}
\right)
\vec{C} = 
\left(
	\begin{array}{c}
	\vdots \\ 0 \\ 0 \\ 0 \\ \vdots
	\end{array}
\right).
\label{null_sp_eq}
\end{equation}
Here $a_s=\mu V_s^2/2\hbar^3\omega$. We solve Eq.~(\ref{null_sp_eq}) for given $a_s$ and 
$a_d$ by using singular value decomposition \cite{nrc} and find the complex energies of the 
quasibound states. 

Due to the structure of the matrix in Eq.~(\ref{null_sp_eq}), if $\varepsilon'$ is a solution 
of Eq.~(\ref{null_sp_eq}), then $\varepsilon'+n$ is also a solution, where $n$ is an integer. 
We limit ourselves to the energy range of $0 < \varepsilon_R\leq 1$, which means the poles 
are meaningful only in the domain $D_0$. This constraint determines the signs of the complex 
square root functions in the diagonal of the matrix in Eq.~(\ref{null_sp_eq}) as described 
in Sec.~II. Using the same notation as the one used in Sec.~II the signs can be expressed by 
$(\cdots, +, +, \ominus, -, -, \cdots)$. 

\begin{figure}
\center
\includegraphics[height=\size cm,angle=0]{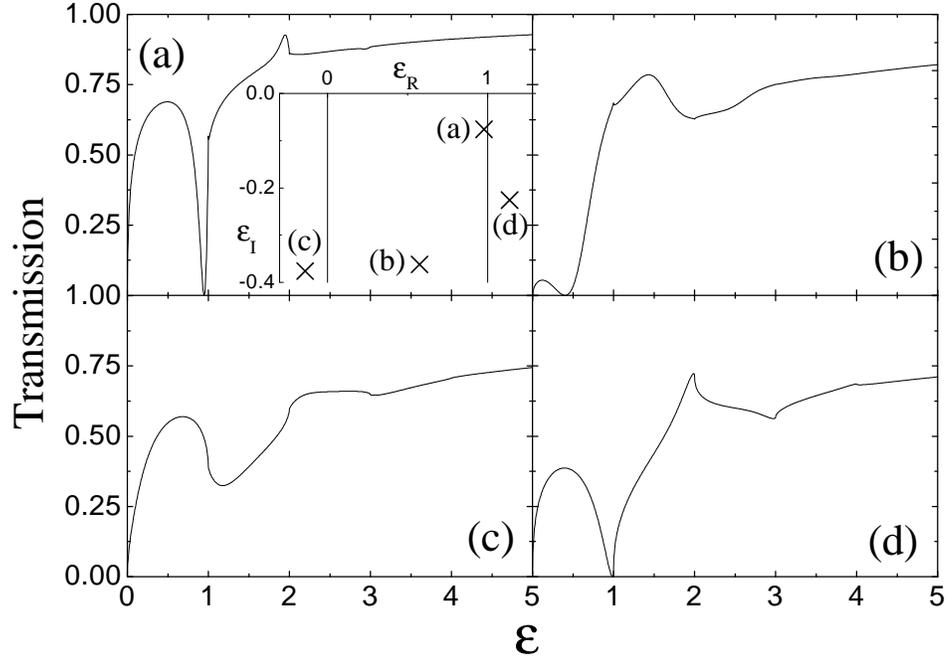}
\caption{Transmission through an oscillating potential for (a) $a_d=0.2$, (b) $a_d=0.6$,
(c) $a_d=1.0$, and (d) $a_d=1.2$ with $a_s=0$, where $a_d$ and $a_s$ represent the strength
of the oscillating potential and of the static part, respectively (for exact definitions see 
the text). The inset in (a) shows the locations of the poles in complex energy plane for each 
case.}
\label{fig2}
\end{figure}

\begin{figure}
\center
\includegraphics[height=\sizebig cm,angle=0]{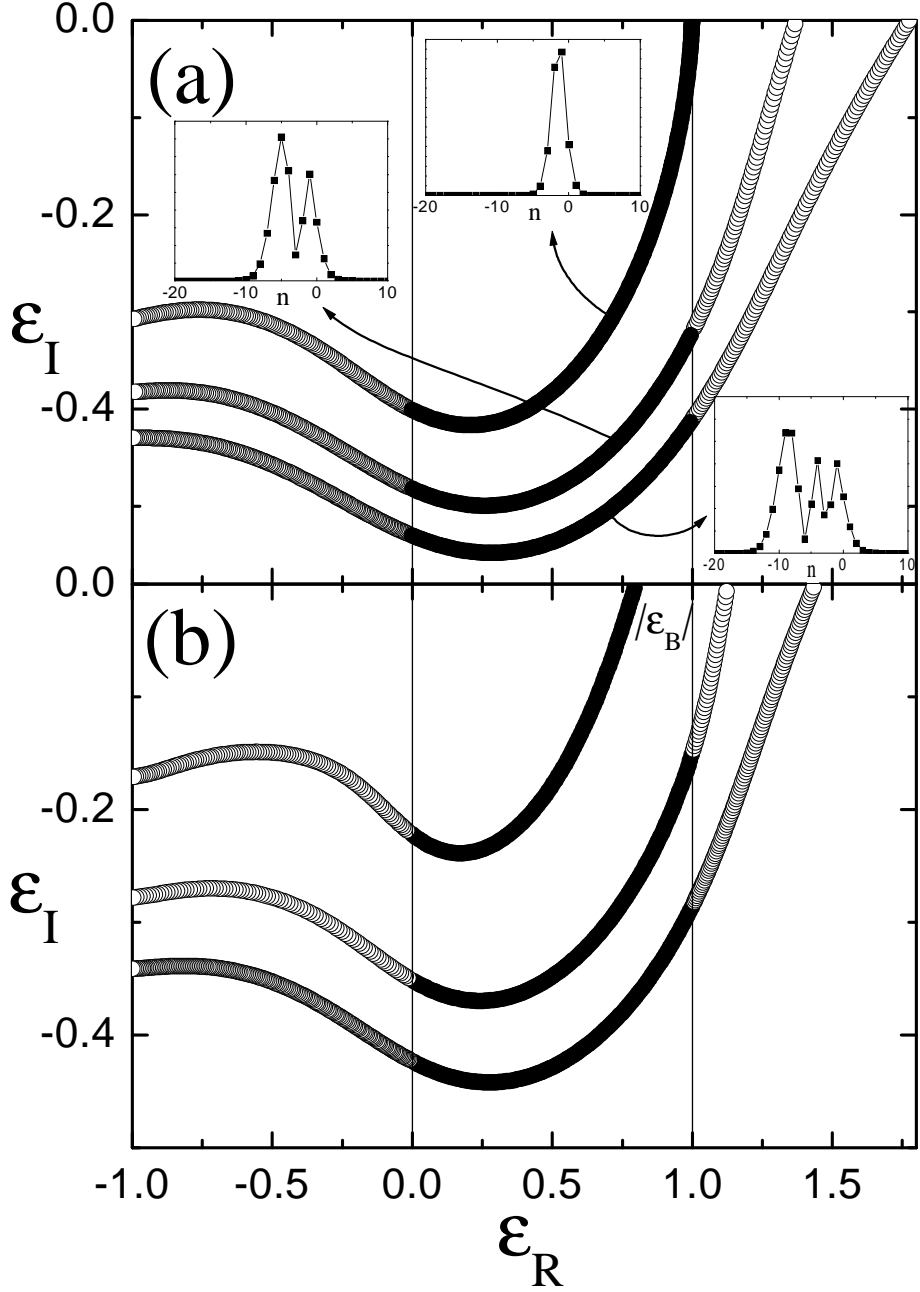}
\caption{(a) Trajectory of a quasibound state (the filled circles) as $a_d$ increases. The 
open circles represent the trajectory of the quasibound state continued into the neighboring 
domains on the proper Riemann sheet of $D_0$. The quasibound state of the third branch from 
the top appears from the real axis at $a_d \simeq 1.869$, and collides with 
$\varepsilon_R = 1$ and $0$ at $a_d \simeq 2.325$ and $2.935$, respectively (see the text 
for the first and the second branches). Inset: absolute squares of the solution vectors
$\vec{C}$ of Eq.~(\ref{null_sp_eq}) at $a_d=0.5$, $1.5$ and $2.5$ from the top, respectively.
(b) Trajectory of quasibound states as $a_d$ increases ($a_s=-0.2$). At $a_d=0$ the quasibound 
state is located at $\varepsilon=0.8$ due to the existence of a bound state, with binding
energy $\varepsilon_B=0.2$. The quasibound state collides with $\varepsilon_R = 0$ at 
$a_d \simeq 0.46$. A second branch of quasibound states appears from the real axis at 
$a_d \simeq 0.525$, and collides with $\varepsilon_R = 1$ and $0$ at $a_d \simeq 0.645$ and 
$1.22$, respectively. Finally, a third branch appears from the real axis at $a_d \simeq 1.221$, 
and collides with $\varepsilon_R = 1$ and $0$ at $a_d \simeq 1.475$ and $2.035$, respectively.}
\label{fig3}
\end{figure}

Figure \ref{fig3}(a) shows a trajectory of the quasibound state as $a_d$ is varied with
$a_s=0$. The quasibound state of the first branch from the top originates from $(1,0)$
at $a_d=0$, moves towards the domain $D_{-1}$, and crosses the domain boundary 
$\varepsilon_R=0$ at $a_d \simeq 0.935$. 
The trajectory of the pole helps to understand the transmission curves in Fig.~\ref{fig2}.
The resonance structure of the transmission follows the pole, as is shown in Fig.~\ref{fig2} 
and the inset of Fig.~\ref{fig2}(a). Since in the system considered here the 
transmission zero always takes place on the left of the pole in the real energy direction, 
it collides with the domain boundary $\varepsilon_R = 1$ prior to the pole and disappears 
first at $a_d \simeq 0.782$. In Ref.\cite{Martinez01} it is concluded that the quasibound 
state vanishes when the transmission zero disappears. Even after the pole moves out of 
the domain $D_0$, however, it still affects the transmission line shape, as shown in 
Fig.~\ref{fig2}(c).
There is no abrupt change of physical quantities, e.g. the 
transmissions, the Friedel phases, and the Wigner delay time (these two will be shown 
below). The transmission zero disappears when it collides with the point $(0,0)$. 
We can trace the transmission zero even into the non-physical region $\varepsilon_R < 0$ 
by calculating $S$ matrix on the proper Riemann sheet of $D_0$, although the unitarity of 
the $S$ matrix is violated. 

A quasibound state of a second branch appears from the real axis at $a_d \simeq 1.032$ and 
moves to the left. It passes through the domain boundaries $\varepsilon_R=1$ at 
$a_d \simeq 1.31$, while the transmission zero appears in advance at $a_d \simeq 1.165$ 
as shown in Fig.~\ref{fig2}(d). The pole finally leaves the domain $D_0$ by passing 
$\varepsilon_R=0$ at $a_d \simeq 1.935$. The second branch of the pole exactly corresponds 
to the case of the schematic shown in Fig.~\ref{fig1}(b). In fact one can find more branches 
of quasibound states in a regular fashion \cite{Martinez01}. It is noted, however, that the 
branches qualitatively differ in 
their solution vectors, i.e., the null space of the matrix of Eq.~(\ref{null_sp_eq}) at the 
energy of quasibound state as shown in the insets of Fig.~\ref{fig3}(a).

Figure \ref{fig3}(b) shows the trajectory of the pole in the case of a delta-function 
impurity with an attractive static potential added to the oscillating potential. A similar 
behavior as in Fig.~\ref{fig3}(a) is observed. The shifted starting point from 
$\varepsilon=1$ at $a_d=0$ is ascribed to the existence of the true bound state of an 
attractive delta function potential with binding energy $E_B=-\mu V_s^2/2\hbar^2$. 
The introduction of 
the static potential makes no effect on the signs of the complex square root functions, i.e.,
the Riemann sheet, which is clear in Eq.~(\ref{null_sp_eq}). 

An interesting difference
takes place in the trajectory of the quasibound states in the oscillating delta function 
impurity when the attractive static potential is present. For small $a_d$ the matrix in 
Eq.~(\ref{null_sp_eq}) can be approximated by a $3\times 3$ matrix involving only $k_{-1}$, 
$k_0$ and $k_1$, which corresponds to single-photon processes (absorption and emission). 
Using the results of Appendix B of Ref.\cite{Bagwell92} it can be seen that this $3\times 3$ 
matrix is not enough for the case without a static potential to describe the shift along 
the real energy axis of the quasibound states as $a_d$ is increased. At least a $5\times 5$ 
matrix is needed to include the shift, which implies two-photon process. The mathematical 
reason is the fact that the slope $d\varepsilon_I/d\varepsilon_R|_{a_d=0}=\infty$ 
in Fig.~\ref{fig3}(a). On the other hand, in the case with an attractive potential 
$3\times 3$ matrix is enough to obtain a shift of the pole along the real energy axis, as 
shown in Fig.~\ref{fig3}(b) because $d\varepsilon_I/d\varepsilon_R|_{a_d=0}$ has a finite 
value. With an attractive potential, a bound state already exists, and the oscillating 
potential leads to an ac-Stark shift \cite{Bakos77} of that energy level. Without a static 
potential, however, one photon is needed to generate a bound state, and additional photons 
are needed to produce the ac-Stark-like shift.


\subsection{Friedel sum rule}

\begin{figure}
\center
\includegraphics[height=\size cm,angle=0]{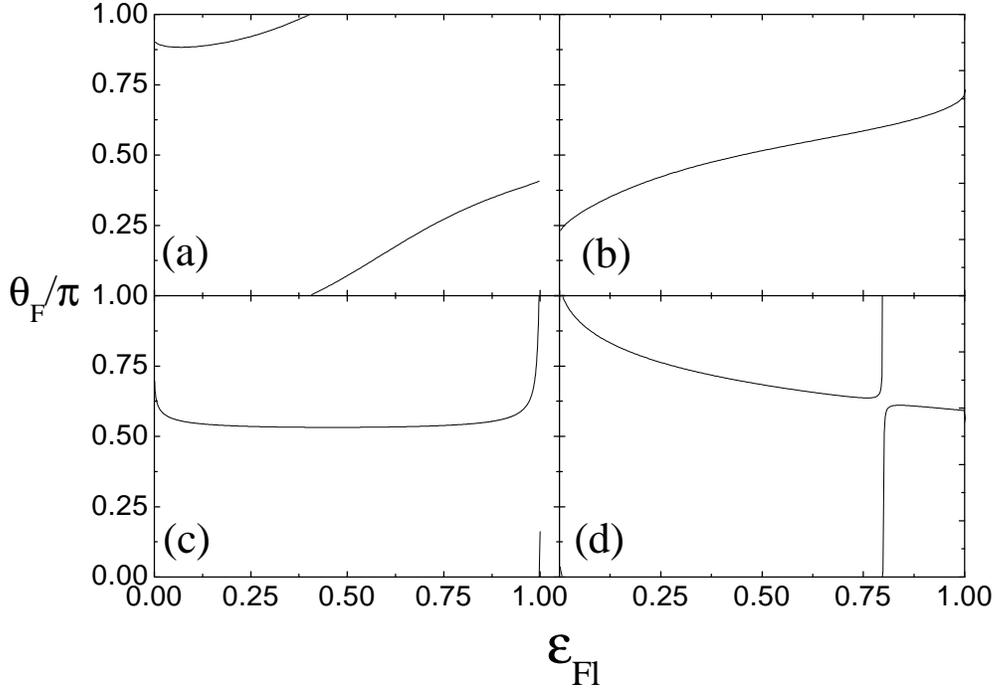}
\caption{Friedel phases as a function of Floquet energy $\varepsilon_{Fl}$ for (a) $a_d=0.6$, 
(b) $a_d=1.0$, and (c) $a_d=0.05$ ($a_s=0$). (d) $a_d=0.001$ with $a_s=-0.2$.}
\label{fig4}
\end{figure}

The number $N_B$ of bound states below an energy $E_f$ (say, the Fermi energy) is related 
to the Friedel phase $\theta_F = \ln Det(S)/2i$  of the scattering matrix $S$ though the 
Friedel sum rule \cite{Friedel58} (see also \cite{Harrison79})
\begin{equation}
N_B=\frac{1}{\pi}\theta_F(E_f).
\end{equation}
This implies that each bound state contributes $\pi$ to the Friedel phase. 
Let us note that the scattering matrix in Eq.~(\ref{osc_smatrix}) is given as a function 
of a Floquet energy $\varepsilon_{Fl}$ ($\equiv E_{Fl}/\hbar\omega$)
and the same holds for the Friedel phase, i.e. it is 
only relevant for $0 \leq \varepsilon < 1$. Figures \ref{fig4}(a) and (b) show Friedel phases 
as a function of $\varepsilon_{Fl}$
for $a_d=0.6$ (the pole is located in $D_0$) and $a_d=1.0$ (the pole is in $D_{-1}$),
respectively, with $a_s=0$. No abrupt change of the Friedel phase is observed at the 
threshold $a_d \simeq 0.782$ of the transmission zero or when the pole crosses the domain 
boundary, $a_d \simeq 0.935$. 

It is noted in Fig.~\ref{fig4}(a) that the Friedel phase does not exhibit a clear $\pi$ shift 
but looks even similar to Fig.~\ref{fig4}(b). We expect the Friedel phase to shift by $\pi$ 
near quasibound states. The absence of a clear $\pi$ shift in Fig.~\ref{fig4}(a) is mainly 
ascribed to the large overlap between the resonance and the neighboring domain since 
the imaginary part of the complex energy of a pole $\varepsilon_I$ is 
quite large, and so is the width of the resonance. Even though $\varepsilon_I$ is quite small 
for very small $a_d$, the overlap is still large enough to violate the $\pi$ phase shift by 
Friedel sum rule as shown in Fig.~\ref{fig4}(c). The quasibound state is located quite close 
to the boundary $\varepsilon_R =1$ because of $d\varepsilon_I/d{\varepsilon_R}|_{a_d=0}=\infty$. 
If we consider the oscillating potential with an additional attractive static potential, 
however, for very small $a_d$ a gap exists between the bound state and the domain boundary 
$\varepsilon_R=1$, which can make the overlap negligibly small, so that the  phase shift $\pi$ 
is clearly seen in Fig.~\ref{fig4}(d). When the Friedel sum rule is applied to the case 
of a multichannel scattering problem, the overlap between the resonance and neighboring domains 
should be carefully considered. 


\subsection{Wigner delay time}

\begin{figure}
\center
\includegraphics[height=\size cm,angle=0]{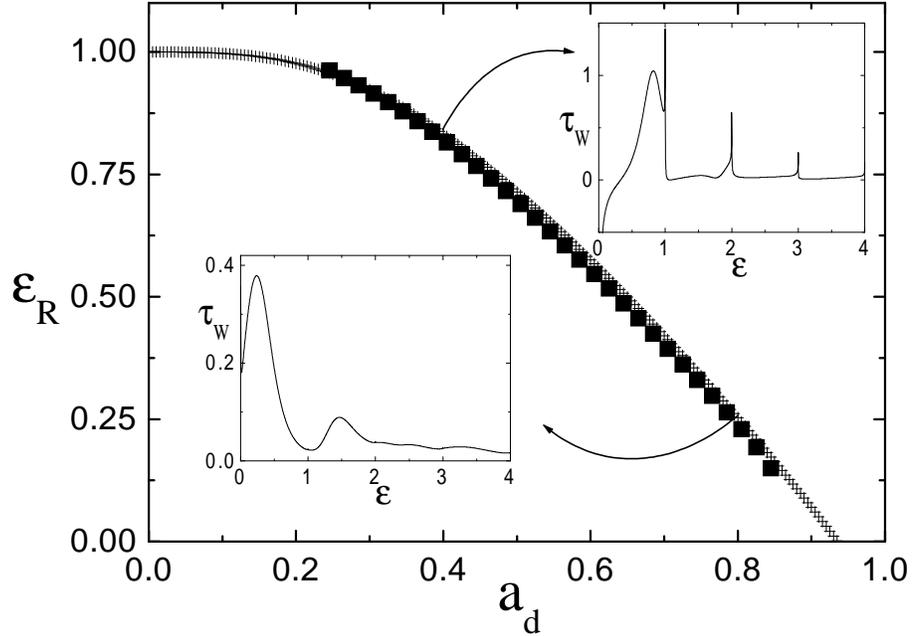}
\caption{The real energies of the trajectory of a quasibound state (crosses) as a function 
of $a_d$ with $a_s=0$, compared to the energies of the local maxima of Wigner delay times
(the filled squares). The insets show Wigner delay times in unit of $1/\omega$ as a function 
of energy $\varepsilon$ for $a_d=0.4$ (left) and $a_d=0.8$ (right).} 
\label{fig5}
\end{figure}

The time scales associated with the quantum scattering have been a quite controversial issue
(see e.g. \cite{Landauer94}) because time is not a Hermitian operator in Hilbert 
space but a parameter. Nevertheless one can define the global time delay of wave-packets 
by using a Hermitian time delay operator (see e.g. \cite{Bracher95}). This time delay operator
turns out to be related to the so-called Wigner-Smith time delay matrix \cite{Wigner55},
whose trace averaged over different channels gives a measure of the time delay of the 
scattered wave caused by the scattering potential (see also \cite{Smith60,Fyodorov97}). 
The Wigner delay time shows a peak at resonance energies, where the maximal value of the 
Wigner delay time approximately corresponds to the decay time $\hbar/\varepsilon_I$. To 
obtain the Wigner delay time we use the eigenvalues of the scattering matrix $S$. Due to 
the unitarity of $S$, all the eigenvalues lie on the unit circle and can be written in 
the form $\exp(i\theta_{\alpha})$. The Wigner delay time is defined by 
\begin{equation}
\tau_W = \hbar \sum_{\alpha}\frac{d\theta_{\alpha}}{dE}|\left< k_n |\theta_{\alpha} \right>|^2,
\end{equation}
where the eigenstate corresponding to the eigenvalue $\theta_{\alpha}$ and an input 
propagating state (or channel) with momentum $k_n$ are denoted by 
$\left| \theta_{\alpha} \right>$ and $\left| k_n \right>$, respectively \cite{Emmanouilidou01}. 
It is worth noting that the Wigner delay time is a function of the energy of the incident particle.
The insets of Fig~\ref{fig5} shows the Wigner delay time as a function of energy of an 
incident particle for two $a_d$'s with $a_s=0$. The peaks of the 
Wigner delay time are located almost exactly at the real part of the complex energy of the 
pole in Fig.~\ref{fig5} (We omitted the data close to the domain boundaries where the Wigner
delay time diverges), under the condition that the pole is located in $D_0$. No abrupt 
change of the Wigner delay time is again observed at the threshold of the transmission zero 
($a_d=0.782$). However, the peak disappears close to the domain boundary $D_{-1}$ since the
pole enters the next domain.


\section{delta-function impurity in quasi-1D wire}

As our second illustrative example we study the simple quasi-1D scattering problem with an
attractive {\em static} delta-function impurity of strength $|\gamma|$. In Ref.\cite{Bagwell90}
it was found that dips in the transmission disappear as $|\gamma|$ is increased. 
Let us again discuss the relation to transmission poles.


\subsection{Scattering matrix formulation}

\begin{figure}
\center
\includegraphics[height=\size cm,angle=0]{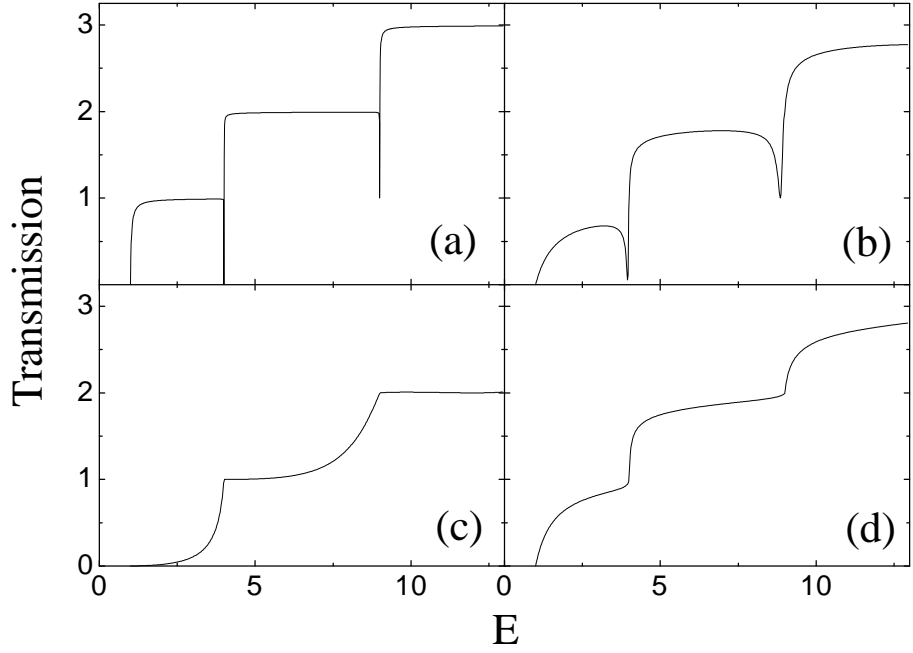}
\caption{Transmission through a delta-function impurity in a quasi-1D wire with (a)
$\gamma=-2.28$ feVcm$^2$, (b) $\gamma=-5.67$ feVcm$^2$, (c) $\gamma=-9.1$ feVcm$^2$, and 
(d) $\gamma=-22.75$ feVcm$^2$, where $\gamma$ represents the strength of the impurity potential.}
\label{fig6}
\end{figure}

The Hamiltonian is given by
\begin{equation}
H = \frac{p^2}{2\mu} + \gamma\delta(x)\delta(y-y_0) + V_c(y),
\end{equation}
where $\gamma <0$, and $V_c$ represents a confinement potential which restricts the movement 
of the particle to the range $0<y<D$ in transversal direction. Using the complete set of the 
transversal modes $\chi(y)=\sqrt{2/D}\sin (n\pi y/D)$, the solution of this Hamiltonian can be 
expressed as
\begin{equation}
\psi_E (x,y) = \sum_{n=1}^{\infty} c_n(x) \chi(y).
\end{equation}
Plugging this expansion into the Schr\"{o}dinger equation and exploiting orthogonality of 
the transversal modes, we obtain the following equation 
\begin{equation} 
\frac{d^2c_m(x)}{dx^2} + k_m^2 c_m(x) = \sum_{n} M_{mn} c_n(x) \delta(x), 
\end{equation} 
where $M_{mn} = (4\mu\gamma/D\hbar^2)\sin(m\pi y_0/D)\sin(n\pi y_0/D)$ and 
$k_n=\sqrt{2\mu E/\hbar^2-(n\pi/D)^2}$. 
Using both the plane waves of Eq.~(\ref{plane_wave}) for $c_n(x)$ and the constraints in 
Eqs.~(\ref{osc_bc1}) and (\ref{osc_bc2}), we obtain the following conditions
\begin{eqnarray}
A_n+B_n = C_n+D_n, \\
ik_n(C_n-D_n)-ik_n(A_n-B_n)=\sum_m M_{nm}(A_m+B_m).
\end{eqnarray}
These can be rewritten as
\begin{equation}
\left(
	\begin{array}{c}
	\vec{B} \\ \vec{C}
	\end{array}
\right)
=
\left(
	\begin{array}{cc}
	(2Q-M)^{-1}M & 2(2Q-M)^{-1}Q \\
	2(2Q-M)^{-1}Q & (2Q-M)^{-1}M
	\end{array}
\right)
\left(
	\begin{array}{c}
	\vec{A} \\ \vec{D}
	\end{array}
\right),
\end{equation}
where $Q_{mn}=ik_n\delta_{mn}$. Following the same procedure as that used in 
Eq.~(\ref{osc_smatrix}) we can obtain the $S$-matrix for this problem and the total 
transmission coefficient
\begin{equation}
T=\sum_{mn({\rm prop.})}\left|t_{mn}\right|^2,
\end{equation}
where the sum extends over all propagating modes. The resulting transmission coefficients 
is shown in Fig.~\ref{fig6}. We use the parameters $D=30$ nm for 
the width of the wire, the mass $\mu=0.067m_e$ as the effective mass of an electron in GaAs, 
$y_0=5D/12$ for the transversal position of the delta-function scatterer, and energy $E$ 
normalized by the $E_1=6.24$ meV. The transmission coefficient depends on the number of 
the propagating channels, which will be discussed in the next subsection.


\subsection{Quasibound states}

Quasibound states can be obtained from the following equation, derived from Green's function 
approach \cite{Boese00},
\begin{equation}
\frac{1}{\gamma}-\sum_n \frac{4\mu}{D\hbar^2} 
\left( \sin\frac{n\pi}{D}y_0 \right)^2\frac{1}{2ik_n} = 0,
\end{equation}
which has to be solved for energy $E_R - iE_I$ [entering via Eq.~(\ref{sqrt})].
It should be mentioned that the location of the quasibound states and the transmission 
coefficients strongly depend on the number of modes $N_{ch}$ \cite{Boese00}. In fact they 
do not converge although $N_{ch}$ is increased. In order to model a realistic impurity
one should take $N_{ch}$ large but finite, and fix it for all numerical calculations. 
$N_{ch}=100$ is chosen in this paper.

\begin{figure}
\center
\includegraphics[height=\size cm,angle=0]{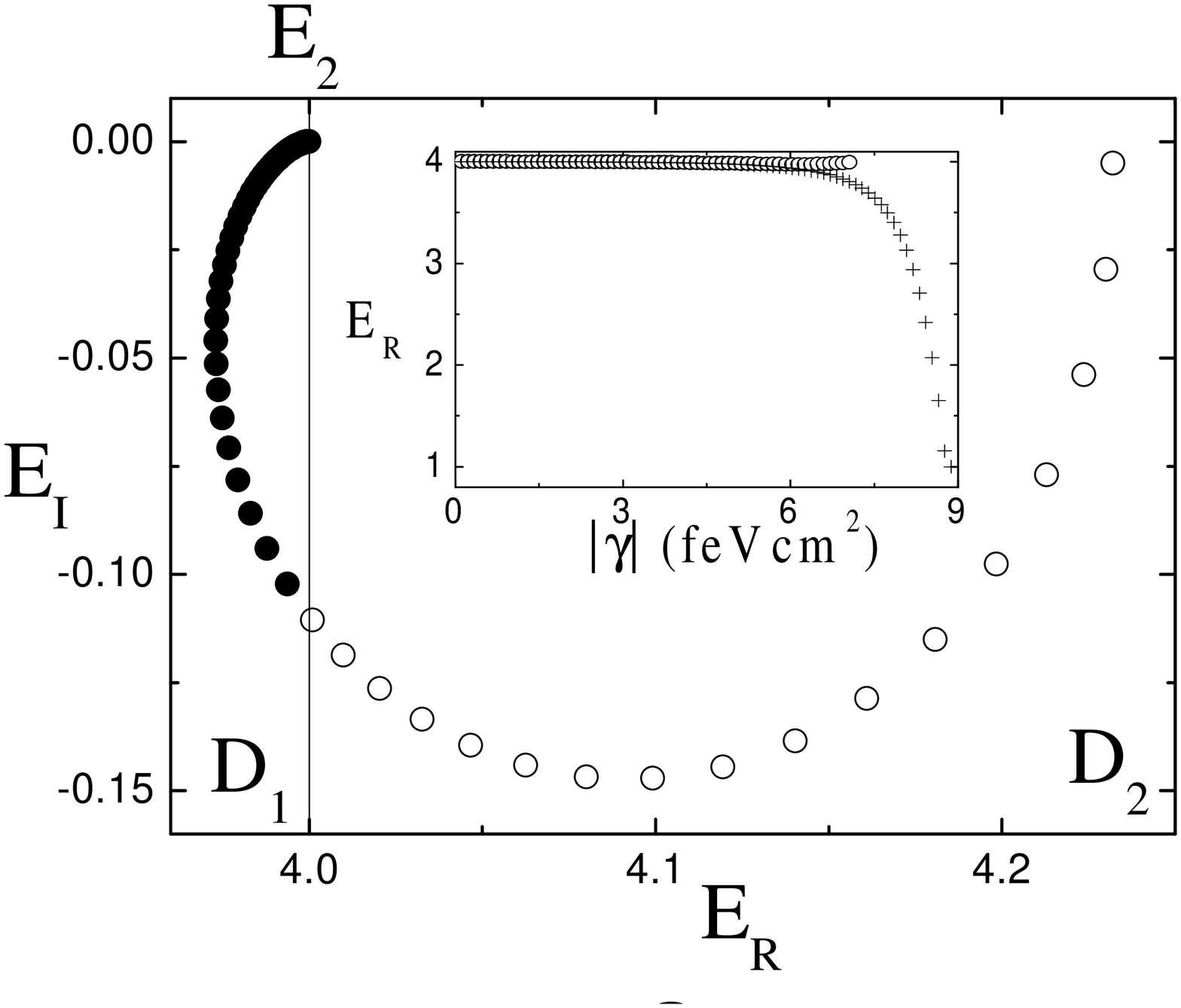}
\caption{Trajectory of a quasibound state for a delta-function impurity in a quasi-1D wire
(filled circles) and its continuation into the neighboring domain. Inset: real energies of 
the trajectory of a quasibound state (open circles) as a function of the strength of the 
delta-function impurity ($\gamma<0$), and the energies of the transmission zeros (crosses).}
\label{fig7}
\end{figure}

Figure \ref{fig7} shows the trajectory of a quasibound state as $\gamma$ is varied. 
The quasibound state crosses the domain boundary $E_2$ at $\gamma \simeq -7.11$ feVcm$^2$,
and enters the next domain $D_2$ (the open circles in Fig.~\ref{fig7}) on the proper Riemann 
sheet of the domain $D_1$, $(\ominus, +, + , \cdots)$. The situation is similar to those out 
of Sec.~III except that the quasibound state collides with the real axis at 
$\gamma \simeq -9.06$ feVcm$^2$ and only one branch of quasibound states exists. 
As $|\gamma|$ increases the transmission zero moves far away from the pole, as is shown in 
the inset of Fig.~\ref{fig7}, and finally vanishes when it collides with the other domain 
boundary $E_1$ at $\gamma = -8.83$ feVcm$^2$.

Apart from the quasibound state, the attractive delta-function impurity also accommodates one
true bound state below $E_1$. Although the quasibound state vanishes at the threshold value 
of $\gamma$, the true bound state exists regardless of the value of $\gamma$, and its energy 
monotonically decreases as $|\gamma|$ increases.


\subsection{Friedel sum rule and Wigner delay time}

\begin{figure}
\center
\includegraphics[height=\size cm,angle=0]{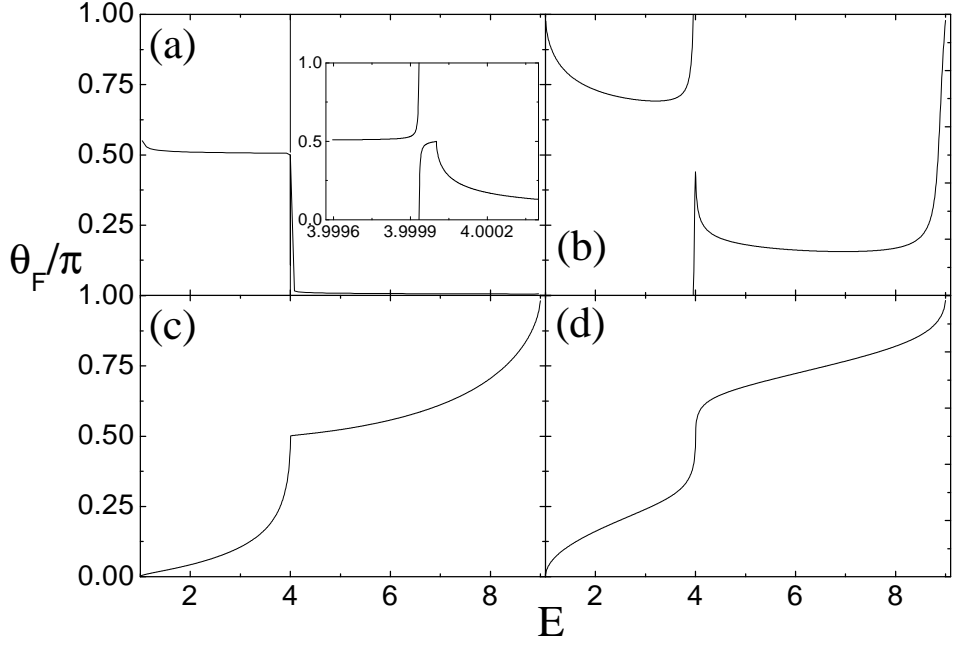}
\caption{Friedel phases in the quasi-1D wire with a static impurity as a function of energy 
$E$ for (a) $\gamma=-0.57$ feVcm$^2$, (b) $\gamma=-5.69$ feVcm$^2$, (c) $\gamma=-9.1$ feVcm$^2$, 
and (d) $\gamma=-11.38$ feVcm$^2$. The inset in panel (a) shows the magnification of the 
resonance region.}
\label{fig8}
\end{figure}

\begin{figure}
\center
\includegraphics[height=\size cm,angle=0]{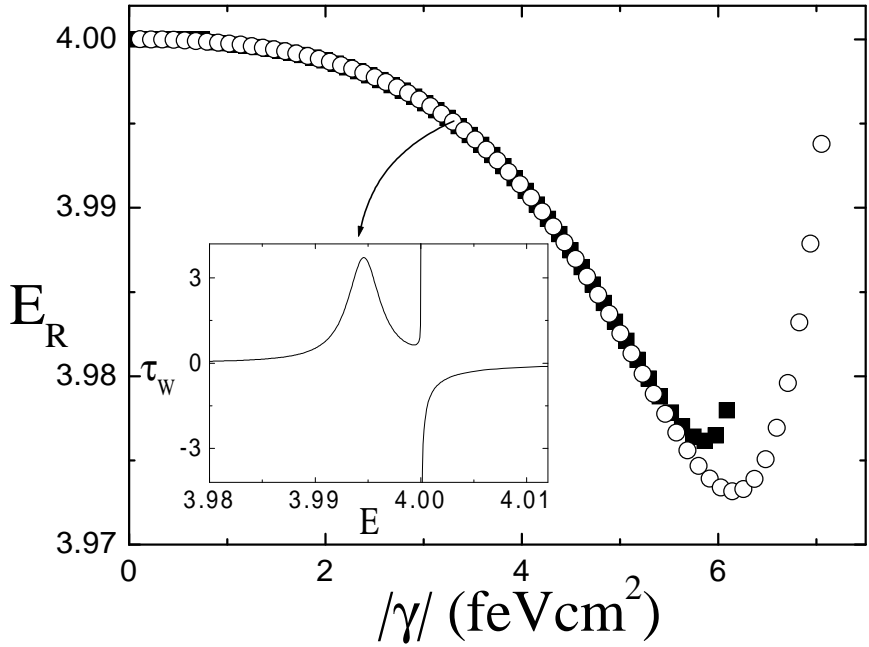}
\caption{The real energies of the trajectory of a quasibound state (open circles) as a 
function of $|\gamma|$, compared to the energies of the local maxima of Wigner delay times 
(filled squares). The inset shows Wigner delay time $\tau_W$ in units of $10^{-12}$ sec as 
a function of energy $E$ for $\gamma=-3.41$ feVcm$^2$, with the local maximum at $E=3.995$.}
\label{fig9}
\end{figure}

The Friedel phase of the impurity in the quasi-1D wire is shown in Fig~\ref{fig8}. The 
$\pi$ phase shift by quasibound state is clearly observed for very small $|\gamma|$, but 
it is hard to be resolved for larger $|\gamma|$ due to the overlap between the resonance 
and the neighboring domain. Figure \ref{fig9} shows the peaks of Wigner delay time and 
the real part of the complex energy of the pole, which coincide very well for $|\gamma| 
\leq 5.5$ feVcm$^2$. For larger $|\gamma|$ it is again difficult to distinguish the peaks 
of the Wigner delay time from the background which diverges at the domain boundary $E_2$
(see Fig.~\ref{fig7}). As $|\gamma|$ 
increases the pole first moves away from $E_2$, and then  moves back toward $E_2$. The peak 
of the Wigner delay time originating from the pole strongly overlaps with the divergent part 
of the background when it changes the direction. Hence, for large $|\gamma|$ the position
of the peak no longer coincides with the energy of the pole. Friedel phase and Wigner delay 
time of the quasi-1D case behave similarly to the case of an oscillating impurity.


\section{summary}

We have investigated the characteristics of transmission resonances, Friedel phases, and
Wigner delay times in multichannel scattering problems focusing on the trajectories of 
quasibound states in complex energy plane as a scattering strength is changed. Subband
thresholds divide the complex energy plane into sub-domains characterized by appropriate 
boundary conditions for the quasibound states. This set of boundary condition is related 
to the choice of the signs of the complex wave number $k_n$ in each subband. A transmission 
resonance (or, a transmission zero) can vanish when it collides with the subband boundary, 
which, however, does not mean that the pole becomes completely irrelevant, as is seen in 
the transmission line shape, the Friedel phase, and the delay time. Even if a quasibound 
state enters the neighboring domain, it can still affect physics at 
the real energies of the domain from which it originated. We illustrate these findings with 
two simple examples: the scattering problem with the 1D oscillating delta-function impurity 
and the static delta-function impurity in quasi-1D wire. 

\section*{Acknowledgments}
We would like to thank Ingrid Rotter for a careful reading of our manuscript and helpful
comments.


\bibliographystyle{prsty}


\end{document}